\documentclass[12pt,a4paper,man,twoside,noextraspace,apacite,floatsintext]{apa6}

\usepackage{amsmath}
\usepackage{apacite}
\usepackage{bm}
\usepackage{enumitem}
\usepackage{epstopdf}
\usepackage{lscape}
\usepackage{multirow}
\usepackage{verbatimbox}

\title{Model Selection of Nested and Non-Nested Item Response Models using Vuong Tests}
\fourauthors{Lennart Schneider}{R. Philip Chalmers}{Rudolf Debelak}{Edgar C. Merkle}
\fouraffiliations{University of T{\"u}bingen, University of Zurich}{York University}{University of Zurich}{University of Missouri}
\shorttitle{Vuong Tests of Item Response Models}
\abstract{In this paper, we apply Vuong's \citeyear{vuo89} general approach of model selection to the comparison of nested and non-nested unidimensional and multidimensional item response theory (IRT) models. Vuong's approach of model selection is useful because it allows for formal statistical tests of both nested and non-nested models. However, only the test of non-nested models has been applied in the context of IRT models to date. After summarizing the statistical theory underlying the tests, we investigate the performance of all three distinct Vuong tests in the context of IRT models using simulation studies and real data. In the non-nested case we observed that the tests can reliably distinguish between the graded response model and the generalized partial credit model. In the nested case, we observed that the tests typically perform as well as or sometimes better than the traditional likelihood ratio test. Based on these results, we argue that Vuong's approach provides a useful set of tools for researchers and practitioners to effectively compare competing nested and non-nested IRT models. \linebreak \hspace*{0,4in} \textit{Keywords:} item response theory, model selection, Vuong test, likelihood ratio test, likelihood inference}
\authornote{This research was supported in part by the Swiss National Science Foundation under Grant 00019\_152548 and the National Science Foundation under Grant 1460719. Portions of the work were presented at the 2018 International Workshop on Psychometric Computing (Psychoco 2018) and the 2018 International Meeting of the Psychometric Society (IMPS 2018). Correspondence should be sent to Lennart Schneider, Department of Psychology, University of T{\"u}bingen, 72076 T{\"u}bingen, Germany. E-mail: lennart.sch@web.de}
\date{}

{ % for the vertical padding

\begin{document}

\maketitle
Item response theory (IRT) consists of a variety of mathematical and statistical models aimed at describing the interaction between unobserved (latent) psychological constructs (traits) and item characteristics. Most commonly, IRT is adopted to understand examinee response behavior to aptitude tests, psychological inventories, ratings scales, and other forms of (typically categorical) response stimuli at the item and composite test score level. As such, a plethora of related, and often competing, IRT models have appeared in the literature for dichotomous and polytomous item response data. For example, regarding polytomous response data, graded response models \cite{sam69}, (generalized) partial credit models \cite{mur92}, sequential response models \cite{tutz90}, and nominal response models \cite{bock72}, have been studied extensively, where each model may be theoretically suitable for a given empirical investigation.

Aside from selecting a suitable IRT model a priori, which in many applications may itself be difficult, the selection of IRT models often consists of comparing ``best fitting'' models among sets of competing models. Best fitting in this context refers to favoring response models based on statistical decision and information theoretic grounds. This is often achieved by either selecting models that provide more statistically likely fit to the data (i.e., that result in relatively small data-model residuals), or by choosing the most parsimonious of the competing models that also explains the data well.

Depending on the nature of the competing IRT models, various statistical tests can be applied to conduct model selection and comparisons. For instance, when models are nested, model selection can be investigated using the traditional likelihood ratio test approach \cite{neyman28,neyman33}, which is sometimes derived from the difference in $G^2$- or $\chi^2$-statistics \cite{bak04,bock81,reckase09,schilling05,thi88}. In the case when models are not nested, model selection can be performed using information criteria, such as Akaike's Information Criterion \cite<AIC;>{akaike74} or Schwarz's Bayesian Information Criterion \cite<BIC;>{schwarz78}, among others. In this paper, however, we recapitulate Vuong's \citeyear{vuo89} general approach of model selection and apply it to the comparison of both nested and non-nested unidimensional and multidimensional IRT models. 

Briefly stated, Vuong's \citeyear{vuo89} theory consists of three distinct statistical tests related to the \emph{distinguishability} and relative model-fit of nested and non-nested models. Vuong tests have been successfully applied in psychometric contexts such as structural equation modeling \cite<SEM;>{levy07,levy11,meryou16}, with \citeA{meryou16} being the first to make full use of Vuong's framework. They specifically allowed for the calculation of all three statistical tests, including those requiring non-standard model output, through software implementations via the R package {\em nonnest2} \cite{nonnest2}. Vuong's framework has also been investigated when comparing mixture distribution models with different numbers of components \cite{greene94,lo01,nylund07}, although it has been noted that the Vuong tests may be problematic when parameters are on the boundary of the parameter space \cite{jeffries03,wilson15}. Recent extensions of Vuong's \citeyear{vuo89} seminal work have also focused on deriving nonparametric test statistics \cite{clarke01,clarke03,clarke07} and overlapping non-nested models. For instance, \citeA{shi15} proposed a simulation based procedure to achieve correct null rejection rates uniformly over all data generating processes, and \citeA{liao16} extended Vuong's work by deriving a new statistical test for the comparison of semi/non-parametric models that retain optimal asymptotic properties.

In the context of IRT models, \citeA{freeman16} recently applied one of the three Vuong tests (namely, the test of non-nested models) to compare compensatory and non-compensatory multidimensional IRT models, concluding that the test proved useful for correctly identifying the data generating model so long as the correlations between latent dimensions is below $0.8$. In this paper, in addition to the test of non-nested models we also consider the two other Vuong tests, one which allows testing the distinguishability of competing non-nested models before evaluating their relative fit to the data. Testing this assumption is in-line with the original work of \citeA{vuo89} because the distribution of the test of non-nested models under the null hypothesis relies on the assumption that the models are first distinguishable. To our knowledge, neither Vuong's test of distinguishability nor Vuong's test of nested models have been investigated in the context of IRT to date. Additionally, we introduce IRT software that provides functionality to conduct Vuong tests.

In the following pages, we provide a brief summary of a selection of popular IRT models, and describe Vuong's \citeyear{vuo89} theory and the three related statistical tests. We then present the results of several Monte Carlo simulation studies, illustrating the properties of these tests when comparing both nested and non-nested IRT models. Next, we apply the Vuong tests to empirical data consisting of an online questionnaire quantifying a ``nerdiness'' construct. Finally, we conclude with a general discussion regarding the utility and future use of the Vuong tests in the context of IRT investigations. To facilitate future applications, we have extended the functionality of the R package {\em nonnest2} \cite{nonnest2} to allow the Vuong tests to be easily conducted on IRT models fitted via the R package {\em mirt} \cite{mirt}.

\section{Theoretical Background}

In this section, we provide background and notation on the \citeA{vuo89} test statistics. Related discussion of the test statistics can also be found in \citeA{levy07} and \citeA{meryou16}.

\subsection{Models and Estimation}

Let $X_{ij}$ be the response from person $i$ ($i=1,\ldots,N$) on item $j$ ($j=1,\ldots,J$), with item $j$ having $K_j$ categories. We consider $M$-dimensional IRT models of the form
\begin{align}
  X_{ij} | \bm{\theta}_i, \bm{\Psi} &\sim \text{Multinomial}(n=1, p_{ij0}, p_{ij1}, \ldots, p_{ij[K_j-1]}), \\
  \label{eq:model}
  \log \left ( \frac{p^*_{ijk}}{1 - p^*_{ijk}} \right ) &= \beta_{jk} + \displaystyle \sum_{m=1}^M \alpha_{jm} \theta_{im}\ \ \ k=0,\ldots,K_j-1,
\end{align}
where $\bm{\theta}_i$ contains person parameters (i.e., factor or trait scores) for person $i$; $\bm{\Psi}$ contains item parameters and person hyper-parameters (e.g., means, variances, covariances); and $p^*_{ijk}$ is a function of the original category probabilities, $p_{ij0}, p_{ij1}, \ldots, p_{ij[K_j-1]}$.

The above equations cover many popular IRT models. For example, the graded response model \cite<GRM;>{sam69} is obtained by setting
\begin{equation}
  p^*_{ijk} = P(X_{ij} \geq k),
\end{equation}
and the generalized partial credit model \cite<GPCM;>{mur92} is obtained by setting
\begin{equation}
  p^*_{ijk} = P(X_{ij} = (k+1) | X_{ij} \in\ \{k, (k+1)\}).
\end{equation}
Further, when $K_j=2$ for all $j$, the models both reduce to $M$-dimensional two-parameter logistic models ($M$d-2PLM). Note that, for multidimensional models, there is a distinction to be made with respect to between-item and within-item multidimensionality; in the former, one restricts each item to only load on one dimension, resulting in a so-called simple structure; in the latter, one allows each item to load on each dimension \cite<see, e.g.,>{adams97}. The usual unidimensional 2PLM results when $M=1$. Finally, Rasch-like versions of the models can be obtained by setting $M=1$, fixing $\alpha_{j1}=1$ for all $j$, and freely estimating the latent variance hyper-parameter. Across all versions of this model, we assume that the $\bm{\theta}_i$ are random variables (typically from a multivariate normal distribution), leading to models estimated via marginal maximum likelihood (marginal ML). However, the test statistics described below are potentially applicable to models estimated via other ML methods, e.g., conditional ML \cite<see>{bak04}.

Focusing on marginal ML, models are estimated by choosing values of $\bm{\Psi}$ to maximize the log-likelihood
\begin{equation}
  \ell(\bm{\Psi}; \bm{x}_1, \dots, \bm{x}_N) ~=~
  \sum_{i = 1}^N \ell(\bm{\Psi}; \bm{x}_i) ~=~
  \sum_{i = 1}^N \log f(\bm{x}_i; \bm{\Psi}),
\end{equation}
where the log-likelihood for person $i$ is marginalized over $\bm{\theta}_i$, i.e.,
\begin{equation}
  \ell(\bm{\Psi}; \bm{x}_i) = \log \int \prod_{j=1}^J f(x_{ij}; \bm{\Psi}, \bm{\theta}) g(\bm{\theta}; \bm{\Psi}) \partial \bm{\theta},
\end{equation}
with $g(\bm{\theta}; \bm{\Psi})$ often following a $\mathcal{N}_{M}(\bm{0},\bm{\Sigma})$ distribution with correlations and covariances as person hyper-parameters. Maximizing the log-likelihood function involves searching for values of $\bm{\Psi}$ such that the gradient of the log-likelihood is $\bm{0}$, and therefore has reached a (locally) optimal parameter set. The gradient can be represented as the sum of {\em scores} across individuals, i.e.,
\begin{equation}
  s(\hat{\bm{\Psi}}; \bm{x}_1, \dots, \bm{x}_N) = \sum_{i=1}^{N} s(\hat{\bm{\Psi}}; \bm{x}_i) ~=~ \bm{0},
\end{equation}
where
\begin{equation}
  \label{eq:score}
  s(\bm{\Psi}; \bm{x}_i) ~=~ \left(
    \frac{\partial \ell(\bm{\Psi}; \bm{x}_i)}{\partial \Psi_1},
    \dots,
    \frac{\partial \ell(\bm{\Psi}; \bm{x}_i)}{\partial \Psi_P}
  \right)^\prime
\end{equation}
contains derivatives of person $i$'s log-likelihood across all $P$ parameters in $\bm{\Psi}$. Computation of these derivatives is aided by an identity attributed to \citeA{lou82}, which is particularly useful when the IRT models are estimated using the Expectation-Maximization (EM) algorithm; \citeA<see>{bak04} and \citeA{gla98} for further detail.

Following estimation, we can obtain standard errors of parameter estimates via computation of the model's observed or expected parameter information matrix, $\bm{I}(\bm{\Psi})$. Unfortunately, these matrices are more complicated to compute for IRT models than for many other types of statistical models, particularly when the EM algorithm is adopted during estimation \cite{bock81}. Recently, however, \citeA{cha18} demonstrated an accurate and efficient numerical scheme to obtain the observed information matrix which capitalize on Oakes' \citeyear{oak99} identity \cite<see also>{pri17}. Throughout this paper, we utilized the observed information matrix results obtained via the Oakes identity approximation method described by \citeA{cha18}.

\subsection{Vuong Statistics}

The test statistics studied in this paper are generally used to compare two models, which we label Model A and Model B. Once the two models are estimated, we have two parameter vectors, $\bm{\Psi}_A$ and $\bm{\Psi}_B$, along with their respective information matrices, $\bm{I}(\bm{\Psi}_A)$ and $\bm{I}(\bm{\Psi}_B)$. Each individual also has a log-likelihood $\ell(\cdot)$ and a score vector $s(\cdot)$ under each model. These are the building blocks used to construct the Vuong test statistics.

\subsubsection{Nesting, Non-nesting, and Equivalence}

Before defining the test statistics, we define different types of relationships between models. Researchers are generally familiar with nested models, whereby one model (a ``reduced model'') is a special case of another model (a ``full model''); that is, the reduced model's predictions are a subset of the full model's predictions. However, researchers are often less familiar with the concept of ``overlapping'' classification of non-nested models. If two non-nested models are overlapping, they make identical predictions in some populations, but not in others. Conversely, in the non-overlapping or strictly non-nested case, two non-nested models make unique predictions in all populations. The ``overlapping'' attribute is somewhat similar to model equivalence, which is often discussed in the context of SEM models \cite<e.g.,>{ben10,hermar13,macweg93}. However, equivalent models make identical predictions across all populations, whereas overlapping models make identical predictions in only some populations. 

To build an intuition for the ``nested'', ``overlapping'' and ``strictly non-nested'' definitions in the context of IRT modeling, consider the following example.
Suppose we administer a psychological inventory consisting of ten dichotomously scored items to a random sample of persons of some population, where the population then is defined by the probabilities of the response vectors \cite<see, e.g.,>{maydeu13}. We could then analyze the data by fitting, e.g., the Rasch model (RM) or the 2PLM. The RM and 2PLM are nested models in that if the 2PLM slopes are restricted to be equal for all items\footnote{Recall, that marginal ML estimation of the RM either requires the restriction of the slopes of all items to be equal or the restriction of all slopes to one and freely estimating the latent variance hyper-parameter.} the model results in the same predictions provided by the RM. In this sense, the probabilities of the response vectors are a subset of the predictions of the 2PLM. Looking at a single item under the RM, the logit transformed probabilities of solving item $j$ given $\theta_i$ (Equation~\ref{eq:model}) are given by a line parallel to the identity (assuming that the slopes are fixed at one) and intercept $b_j$, whereas under the 2PLM, the logit transformed probabilities of solving item $j$ given $\theta_i$ are given by all lines parameterized with slope $a_j$ (which can now vary freely) and intercept $b_j$. Regarding overlapping non-nested models, consider two different restricted 2PLMs, with the first model restricting the slopes of the first five items to be equal and fixing the remaining slopes to one, whereas the second model fixes the slopes of the first five items at one and restricts the slopes of the remaining items to be equal. These models are non-nested and generally make different predictions, but they cannot be distinguished in populations for which the probabilities of the response vectors are based on slopes of one for all items (i.e., the two models overlap). Finally, regarding strictly non-nested models, consider for example a 2PLM that is to be compared to a two-parametric normal ogive model \cite<see, e.g.,>{bock70}.

For pairs of non-nested models, the overlapping concept potentially leads to two separate statistical tests. First, if models are overlapping (or if we are unsure about whether they are overlapping), we can test whether the model predictions are identical in the population of interest; this is a test of {\em distinguishability}. Stated differently, we examine the fit of two models to sample data (which generally will not be identical), and test whether the sample fit statistics could have arisen from models that provide identical fit in the population of interest. If the test indicates indistinguishable models, then we have no basis for choosing one model over the other. However, if the test indicates distinguishable models, we can further examine whether one model provides a ``significantly better'' fit than the other. This second test is akin to the traditional likelihood ratio test, except that the two candidate models are non-nested. Note that this is the test that \citeA{freeman16} investigated.

For pairs of nested models, the distinguishability and likelihood ratio tests can still be carried out to test the same hypotheses as the traditional likelihood ratio test. However, unlike the traditional likelihood ratio test \cite<see, e.g.,>{chusha09,stesha85}, the Vuong test statistics make no assumptions related to the full model being ``correctly specified'' (i.e., the full model potentially may contain the true conditional distribution of the data). This point is further discussed in the next section.

\subsubsection{Statistics}

The Vuong statistics' derivations focus on the Kullback-Leibler (K-L) distance \cite{kullei51} between each model and the population generating model (PGM). A better-fitting model is one whose distance to the PGM is smaller, and two models fit equally well if their distances are equal. The statistics focus on the case-wise log-likelihoods of the fitted models; each observation in the data will have a log-likelihood value under both candidate models. If two overlapping non-nested models are indistinguishable from one another then each observation's log-likelihood will be nearly identical under both models. This concept is tested by computing the variance of differences between log-likelihoods under the two models. Similarly, if two distinguishable non-nested models have the same overall goodness of fit then the mean log-likelihood across observations will be the same for both models. This concept is tested by computing the mean difference between log-likelihoods. 

\paragraph{Test of distinguishability}
Define a population variance in case-wise log-likelihoods as
\begin{equation}
  \label{eq:dist}
  \omega^2_* = \mathrm{VAR} \left [ \log \frac{f_A(\bm{x}_i; \bm{\Psi}^*_A)}{f_B(\bm{x}_i; \bm{\Psi}^*_B)} \right ],
\end{equation}
where $\bm{\Psi}^*_A$ is the Model A parameter vector that is closest to the PGM in K-L distance across the entire population (i.e., where $i$ includes all members of the population), the vector $\bm{\Psi}^*_B$ is defined similarly, and $f_A(\bm{x}_i; \bm{\Psi}^*_A)$ and $f_B(\bm{x}_i; \bm{\Psi}^*_B)$ are the probability density functions of the response vector $\bm{x}_i$ under the respective model. We can formally test the hypothesis that non-nested models are indistinguishable via
\begin{align}
  \label{eq:h0om2}
  H_0\colon&\ \omega^2_* = 0\\
  \label{eq:h1om2}
  H_1\colon&\ \omega^2_* > 0,
\end{align}
with the associated estimate of $\omega^2_*$ being
\begin{equation}
  \label{eq:sampom}
  \hat{\omega}^2_* = \frac{1}{N} \displaystyle \sum_{i=1}^N \left [
    \log \frac{f_A(\bm{x}_i; \hat{\bm{\Psi}}_A)}{
      f_B(\bm{x}_i; \hat{\bm{\Psi}}_B)} \right ]^2 - \left [
    \frac{1}{N} \displaystyle \sum_{i=1}^N \log
    \frac{f_A(\bm{x}_i; \hat{\bm{\Psi}}_A)}{
      f_B(\bm{x}_i; \hat{\bm{\Psi}}_B)} \right ]^2.
\end{equation}
Under~\eqref{eq:h0om2}, Vuong showed that $N \hat{\omega}^2_*$ follows a weighted sum of $\chi^2$ distributions, where the weights are computed by taking squared eigenvalues of a matrix that involves the two models' scores and information matrices \cite<see the appendix of>[for technical detail]{meryou16}. Computations involving weighted sums of $\chi^2$ distributions are generally complicated, and the computations are facilitated herein via use of the R package {\em CompQuadForm} \cite{duclaf10}.

\paragraph{Goodness of fit}
Assuming that the two non-nested models are distinguishable, we can proceed to test their fits by comparing the mean log-likelihood under each model. The hypotheses are specified via
\begin{align}
    \label{eq:h0lrt}
    H_0\colon&\ \mathrm{E}[\ell(\bm{\Psi}^*_A; \bm{x}_i)] =
    \mathrm{E}[\ell(\bm{\Psi}^*_B; \bm{x}_i)] \\
    \label{eq:h1lrt}
    H_{1}\colon&\ \mathrm{E}[\ell(\bm{\Psi}^*_A; \bm{x}_i)] \neq
    \mathrm{E}[\ell(\bm{\Psi}^*_B; \bm{x}_i)],
\end{align}
where the direction of $H_1$ is typically considered in drawing final conclusions (i.e., instead of concluding that the two models differ in fit, the researcher interprets one model as fitting better than the other).

The test statistic associated with these hypotheses is similar to a paired-samples $t$-test: An observation has a log-likelihood under each model, and the test statistic is based on the mean and variance of differences between log-likelihoods across observations. Formally, the test statistic is
\begin{equation}
  \label{eq:zlrt}
  \mathrm{LR}_{AB} = N^{-1/2} \displaystyle \sum_{i=1}^N \log \frac{f_A(\bm{x}_i;
    \hat{\bm{\Psi}}_A)}{f_B(\bm{x}_i; \hat{\bm{\Psi}}_B)},
\end{equation}
which, under~\eqref{eq:h0lrt}, converges in distribution to $\mathcal{N}(0, \omega^2_*)$ when models are distinguishable \cite[Theorem 5.1]{vuo89}. Note that in {\em nonnest2} \cite{nonnest2}, this test statistic is rescaled, resulting in a $Z$ test statistic following the standard normal distribution under the null hypothesis.

\paragraph{Testing nested models}
In the case of nested models, the two statistics described above (Equations~\ref{eq:dist} and~\ref{eq:zlrt}) are alternative ways of testing the same hypothesis. Assuming that Model B is nested within Model A\footnote{For a rather formal but mathematically precise definition of nestedness of conditional models, the reader is referred to Definition 4 and Assumption A8 of \citeA{vuo89}.}, the hypothesis that the restrictive Model B fits as well as the less restrictive Model A, and the alternative hypothesis that the less restrictive Model A provides a better fit than Model B, can be written as
\begin{align}
  \label{eq:tradh0}
  H_0\colon&\ \bm{\Psi}_A \in\ h(\bm{\Psi}_B) \\
  \label{eq:tradh1}
  H_1\colon&\ \bm{\Psi}_A \not\in\ h(\bm{\Psi}_B),
\end{align}
where $h(\cdot)$ is a function translating the $M_B$ parameter vector to an equivalent $M_A$ parameter vector.

The limiting distribution of the test statistics depends on whether or not we assume that Model A is correctly specified. If we do make this assumption (which is commonly employed for traditional tests of nested models) then both statistics ($N \hat{\omega}^2_*$ and $\mathrm{LR} = 2 N^{1/2} \mathrm{LR}_{AB})$ weakly converge to the usual $\chi^2$ distribution. If we do not make this assumption then the statistics strongly converge to weighted sums of $\chi^2$ distributions, where the weights again involve the eigenvalues of a matrix containing the models' scores and information matrices.

The test statistics described above are implemented for many classes of models in the R package {\em nonnest2} \cite{nonnest2}. As part of the current paper, the package functionality was extended to IRT models estimated via the R package {\em mirt} \cite{mirt}, which is often used to obtain parameter estimates in IRT models using the marginal ML criteria. It should be noted that Vuong's theory can also be used for the computation of AIC and BIC confidence intervals for non-nested models \cite<see, e.g.,>{meryou16}, which are not discussed herein. We use these package extensions throughout the paper to study and illustrate the Vuong test statistics' applications to IRT.

\subsection{Alternative Methods}

We now briefly discuss some widely used methods that aim at similar model comparison and decision goals. Later, we will compare these methods against Vuong's tests through a selection of simulation studies.

\subsubsection{Nested Models and Likelihood Ratio Tests}

Several authors have discussed the use of likelihood ratio tests for comparing the relative model fit of two nested models, both within the context of IRT \cite<e.g.,>{reckase09} and in factor analysis \cite<e.g.,>{hayashi07}. The traditional likelihood ratio test \cite{neyman28,neyman33}, and the related difference in $G^2$ or $\chi^2$, have been shown to follow an asymptotic $\chi^2$ distribution under the null hypothesis under a wide range of conditions and models; examples include log-linear models \cite{haberman77} and factor analytic models \cite{amemiya90}. \citeA{drton09} discusses some factor analytic models for which the limiting distribution of the traditional likelihood ratio test can be proven to be no longer $\chi^2$ (i.e, when testing the complete independence model against the one-factor model). A typical application of likelihood ratio tests is the assessment of the latent trait dimensionality of a dataset \cite<e.g.,>{maydeu06, schilling05, tollenaar03}. In the context of IRT, \citeA[Chapter 7.2.4]{reckase09} discusses the use of differences in $\chi^2$ for determining the number of dimensions of IRT models, and states that this procedure overestimates the true number of dimensions (i.e., results in inflated Type I error rates). In contrast, \citeA{tate03} found this procedure to generally work well when slopes were restricted to one (e.g., between-item multidimensional Rasch-like models).

In the context of (exploratory) multidimensional IRT, a typical parametrization of $M$-dimensional models consists of freely estimating the (item) slopes of all $M$ dimensions (except for one item slope for the second dimension, two item slopes for the third, three item slopes for the fourth, and so on, which are fixed at zero to resolve the rotational indeterminacy of the model) and assuming the $M$ latent dimensions to have means of zero and the identity matrix as the covariance matrix. When the null hypothesis of the traditional likelihood ratio test holds (i.e., data follow the $M-1$ dimensional model) the $M$ dimensional model can be reduced to the $M-1$ dimensional model by restricting the (item) slopes of the $M$-th dimension to zero. Looking at the $M$-th latent dimension, this would also imply a latent variance of zero; however, this latent variance is upwardly biased due to the parametrization of the variance covariance matrix. Overall, one could argue that this results in a misspecification scenario and the traditional likelihood ratio test should not be used to begin with. To our knowledge, this has not been explicitly discussed in the literature of IRT as of yet. However, in the literature of exploratory factor analysis, \citeA{hayashi07} discuss that, when the number of factors being modeled exceeds the true number of factors, the traditional likelihood ratio test may no longer follow a $\chi^2$ distribution due to rank deficiency and non-identifiability of model parameters.

Finally, as it has already been stated, a basic assumption of the use of likelihood ratio tests for evaluating the relative model fit is that the less restrictive of the two nested models is correctly specified. If this assumption is not met, $p$-values are in general no longer uniformly distributed under the null hypothesis. Simulation studies in IRT and factor analysis indicate that the asymptotic distribution is no longer $\chi^2$ if neither of the two models being compared is the true model \cite{maydeu06, yuan04}. In contrast, the asymptotic distribution of Vuong's test statistics are not based on the assumption that one of the two competing models is the true model. We therefore expect the Vuong tests to sometimes exhibit different behavior than the traditional likelihood ratio test, especially in the case of comparing nested models of different dimensions, and when neither of the two competing models is correctly specified.

\subsubsection{Information Criteria}

Information criteria (e.g., AIC and BIC) are widely known and commonly used tools for model selection. Taking into account the fit and complexity of the competing models, information criteria aim at providing an index of model fit, where a lower index expresses a better model fit in terms of both data-model fit as well as parsimony; for an extensive overview \citeA<see>{burnham02}. While the application of information criteria is not limited to nested models, explicitly choosing a ``better fitting'' model can be somewhat difficult. Popular approaches include ``rules of thumbs'', such as observing an absolute difference in AIC larger than ten suggests ``strong'' support for the model with the lower AIC \cite{burnham04}, or simpler approaches such as selecting the model with the lowest index regardless of the absolute difference. \citeA{kang07} and \citeA{kang09} studied the performance of the AIC and BIC in the context of model selection of both dichotomous and polytomous IRT models, among other Bayesian measures of fit and other test statistics, and found them to generally perform well in that they often indicated correct preference for the true PGM.

\subsubsection{Assessing Absolute Model Fit}

In the analysis of categorical data, $G^2$, and the related $\chi^2$-statistic, are used to assess the absolute fit of specific models \cite{agresti02}. For testing this hypothesis in IRT, limited information fit statistics have been proposed as an alternative to account for the sparsity of the underlying contingency table. Two prominent examples in IRT are $M_2$ \cite{maydeu05} and $M_2^*$ \cite{cai13}. These statistics aim at testing a different null hypothesis than Vuong's test statistics in that they compare the fit of a given model against the first and second moments of the data. Under the null hypothesis, the $M_2$ statistic and its variants are asymptotically $\chi^2$ distributed with degrees of freedom equal to the total number of multivariate moments used for testing minus the number of model parameters estimated \cite<see, e.g.,>{maydeu05}. Finally, a hybrid variant of the $M_2$ statistics exists known as the $C_2$ statistic \cite{cai14}, where only the bivariate moments are collapsed. The $C_2$ statistic is useful when fitting polytomous models to shorter tests that do not have sufficient degrees of freedom to compute $M_2^*$ and contain data tables that are too sparse to effectively compute $M_2$.
\section{Simulation 1: Non-Nested Models}

In this and the following sections, we study the Vuong tests' application to IRT using both simulations and real data. We also compare Vuong's tests to the aforementioned model assessment approaches discussed above to evaluate how effective the Vuong tests are relative to previously studied popular methods.

In Simulation 1.1, we compare the fit of two non-nested 2PLMs (as introduced in the theoretical background of nesting, non-nesting and equivalence), which restrict different item slopes to one. Due to the data generating process being the RM, these two models are theoretically indistinguishable; hence, the test results should not demonstrate any systematic preferences for or against a given model. Similarly, in Simulation A.1 (available in the appendix), we compare the fit of the GRM to the GPCM when the data generating process does not follow either model, but rather data is generated from an ``uninformative'' binomial distribution. Note that contrary to Simulation 1.1, this does not necessarily imply that the null hypothesis of Vuong's test of distinguishability holds. Finally, in Simulation 1.2, we compare the fit of the GRM to the GPCM with the data generated under a hybrid model, where items follow either of the two competing models.

In all three simulations, we study Vuong's test of distinguishability, as well as Vuong's test of non-nested models, and compare these to the AIC and the $M_2$ or $M_2^*$ statistics, where applicable.\footnote{We do not include the BIC for further comparison because the models share the same number of estimated model parameters.} All models in this section were estimated via marginal ML, assuming the person parameters follow the standard normal distribution. The EM algorithm for marginal ML estimation was implemented using the default estimation criteria found in {\em mirt} \cite{mirt}, with the exception that up to $5000$ EM cycles were allowed before the algorithm was terminated; otherwise, the algorithm was terminated early (i.e., ``converged'') if all elements of the sets of estimates between two successive EM cycles fell below $|0.0001|$. Simulation results are reported based on the replications in which both models converged. Test statistics were evaluated at an $\alpha$ of $0.05$.

\subsection{Simulation 1.1: Comparing Non-Nested 2PLMs}

\subsubsection{Method}

Simulation conditions were defined by the number of persons, $N = 500$, $1000$, or $2000$, and the length of the test, $J = 10, 20, 30$, or $40$. Data were generated under the RM, where intercepts and person parameters were drawn from the standard normal distribution. In each condition we generated 1000 datasets and on each dataset two non-nested 2PLMs were fit. The first 2PLM restricted the slopes of the first half of the items to be equal while restricting the slopes of the second half to one. The second 2PLM restricted the slopes of the second half of the items to be equal while restricting the slopes of the first half to one. After fitting the models we computed four statistics: Vuong's test of distinguishability, Vuong's test of non-nested models, and each model's AIC and $M_2$ statistic. We checked whether the models could be distinguished, and if this was the case, whether the non-nested test implied preference of one model over the other one. We further checked which model was to be preferred based on the lower AIC, and whether the $M_2$ statistic indicated bad model fit.

\subsubsection{Results}

\begin{table}
\footnotesize
\centering
\addvbuffer[12pt]{
\begin{threeparttable}
\caption{\label{tab:sim_11} Simulation 1.1: Comparing two Non-Nested 2PLMs When Data Follow the RM.}
\begin{tabular}{l c c c c c c c c c c c}
\hline
{} & {} & \multicolumn{10}{c}{Empirical Preference/Rejection Rates} \\ \cline{3-12}
{} & {} & {} & \multicolumn{4}{c}{2PLM\textsubscript{1}} & {} & \multicolumn{4}{c}{2PLM\textsubscript{2}} \\ \cline{4-7} \cline{9-12}
{} & {} & {} & \multicolumn{2}{c}{$\mathrm{LRT_{v}}$} & {} & {} & {} & \multicolumn{2}{c}{$\mathrm{LRT_{v}}$} & {} & {} \\ \cline{4-5} \cline{9-10}
$N$ & $J$ & Dist & all & (Dist sgn.) & AIC & $M_2$ & {} & all & (Dist sgn.) & AIC & $M_2$ \\
\hline
500 & 10 & 0.05 & 0.00 & (0.00) & 0.50 & 0.05 &  & 0.00 & (0.00) & 0.50 & 0.06 \\
500 & 20 & 0.05 & 0.00 & (0.00) & 0.50 & 0.06 &  & 0.00 & (0.00) & 0.50 & 0.05 \\
500 & 30 & 0.06 & 0.00 & (0.02) & 0.47 & 0.04 &  & 0.00 & (0.00) & 0.53 & 0.05 \\
500 & 40 & 0.04 & 0.00 & (0.00) & 0.48 & 0.05 &  & 0.00 & (0.00) & 0.52 & 0.04 \\
\hline
1000 & 10 & 0.06 & 0.00 & (0.02) & 0.52 & 0.04 &  & 0.00 & (0.00) & 0.48 & 0.04 \\
1000 & 20 & 0.06 & 0.00 & (0.00) & 0.51 & 0.07 &  & 0.00 & (0.00) & 0.49 & 0.07 \\
1000 & 30 & 0.05 & 0.00 & (0.00) & 0.49 & 0.06 &  & 0.00 & (0.02) & 0.51 & 0.05 \\
1000 & 40 & 0.04 & 0.00 & (0.00) & 0.51 & 0.06 &  & 0.00 & (0.02) & 0.49 & 0.06 \\
\hline
2000 & 10 & 0.04 & 0.00 & (0.00) & 0.49 & 0.05 &  & 0.00 & (0.00) & 0.51 & 0.05 \\
2000 & 20 & 0.06 & 0.00 & (0.02) & 0.49 & 0.06 &  & 0.00 & (0.02) & 0.51 & 0.06 \\
2000 & 30 & 0.05 & 0.00 & (0.00) & 0.51 & 0.05 &  & 0.00 & (0.00) & 0.49 & 0.05 \\
2000 & 40 & 0.05 & 0.00 & (0.00) & 0.52 & 0.04 &  & 0.00 & (0.00) & 0.48 & 0.05 \\
\hline
\end{tabular}
\begin{tablenotes}[flushleft]
{\small
\textit{Note.} all = using all replications for checking the preference of the non-nested Vuong test ($\mathrm{LRT_{v}}$). Dist sgn. = using only the replications in which Vuong's test of distinguishability (Dist) yielded significant results. $N = $ number of persons. $J = $ number of items.
}
\end{tablenotes}
\end{threeparttable}
}
\end{table}

In all conditions and all replications, the EM algorithm converged for both models. Moreover, in all conditions and all replications, second-order tests based on the condition number of the estimated information matrices of the models indicated that possible local maxima were found.

Table~\ref{tab:sim_11} summarizes the simulation results. Regardless of the number of persons and the test length, Vuong's test of distinguishability (Dist) indicates at a nominal Type I error rate of around $5\%$ that the two 2PLMs can be distinguished. Recall that this statistical test is not designed to determine which of the two competing models provides the better fit to the data. Vuong's test of non-nested models ($\mathrm{LRT_{v}}$) almost never indicates preference of one of the two 2PLMs over the other one. However, as outlined in the Introduction section, this test can only be applied validly if the test of distinguishability yielded a significant result beforehand. Looking only at these few replications, the first 2PLM is to be preferred over the second one at a maximum rate of $2\%$, and the second 2PLM is to be preferred over the first one at a maximum rate of $2\%$. As to be expected, performing model selection based on the lower AIC results in choosing either model at a rate of $50\%$. Finally, the $M_2$ statistic indicates bad model fit for either model at a maximum rate of $7\%$.

\begin{figure}
\centering
\includegraphics[width=3in, height=2.4in]{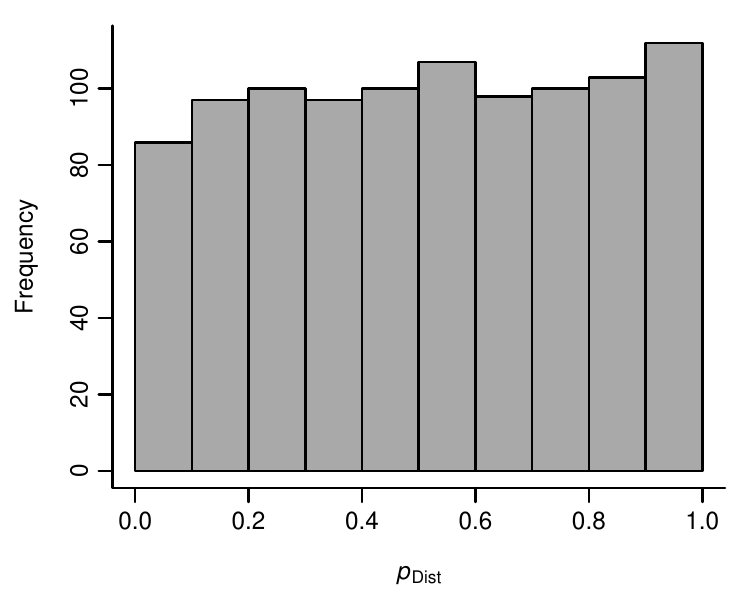}
\caption{\label{fig:sim_11} Simulation 1.1: Histogram of $p$-values for Vuong's test of distinguishability comparing non-nested 2PLMs which are indistinguishable when data follow the RM. $N = 2000$ persons. $J = 10$ items.}
\end{figure}

In this simulation, we further investigated whether the empirical distribution of Vuong's test of distinguishability matches its theoretical distribution under the null hypothesis when comparing non-nested models. We therefore investigated whether the $p$-values are distributed uniformly under the null hypothesis. Figure~\ref{fig:sim_11} shows a histogram of $p$-values for Vuong's test of distinguishability for the scenario of $N = 2000$ and $J = 10$. Looking at this histogram, $p$-values seem to be uniformly distributed. We did not further include this histogram for Vuong's test of non-nested models as only very few replications resulted in the two models being distinguishable, which is a prerequisite to validly conduct Vuong's test of non-nested models.

\subsubsection{Discussion}

In Simulation 1.1, we showed that Vuong's test of distinguishability holds its nominal Type I error rate when comparing non-nested models that are indistinguishable. We have also seen that, under the null hypothesis, Vuong's test of non-nested models shows conservative error control behavior. Based on selecting the model with the lower AIC, however, we observed that choosing either of the competing models occurs at a rate of $50\%$. Finally, the $M_2$ statistic also holds it nominal Type I error rate. In Simulation A.1 (which can be inspected in the appendix), we further expand on the differences in AIC when comparing non-nested models (i.e., the GRM and GPCM). 

\subsection{Simulation 1.2: Data Generated Under a Hybrid Model}

\subsubsection{Method}

In this simulation we investigate the power of the Vuong tests when comparing the GRM and GPCM with the items of the data generating hybrid model following either competing model in varying numbers. Simulation conditions were defined by the number of persons, $N = 500$, $1000$, or $2000$, the length of the test (fixed at $J = 10$), and the number of items of the data generating hybrid model following a four category GPCM ($D = 0, 1, \ldots, 9, 10$). If an item was not generated under the GPCM then it was generated according to a four category GRM. Under both models, slopes were drawn from a log-normal distribution with a mean of zero and a standard deviation of $0.25$. Intercepts were generated based on the distance vector $(1, 0, -1)^\prime$, where for each item this vector was shifted by a random deviance term drawn from the standard normal distribution.

In each condition, we generated 1000 datasets and for each generated dataset we computed four statistics after fitting the models: Vuong's test of distinguishability, Vuong's test of non-nested models, and each model's AIC and $M_2^*$ statistic. We checked whether the models could be distinguished, and if this was the case, whether Vuong's test of non-nested models implied preference of one model over the other. We further checked which model was to be preferred based on the lower AIC, and whether the $M_2^*$ statistic indicated bad model fit.

\subsubsection{Results}

\begin{figure}
\centering
\includegraphics[width=4.5in, height=2.4in]{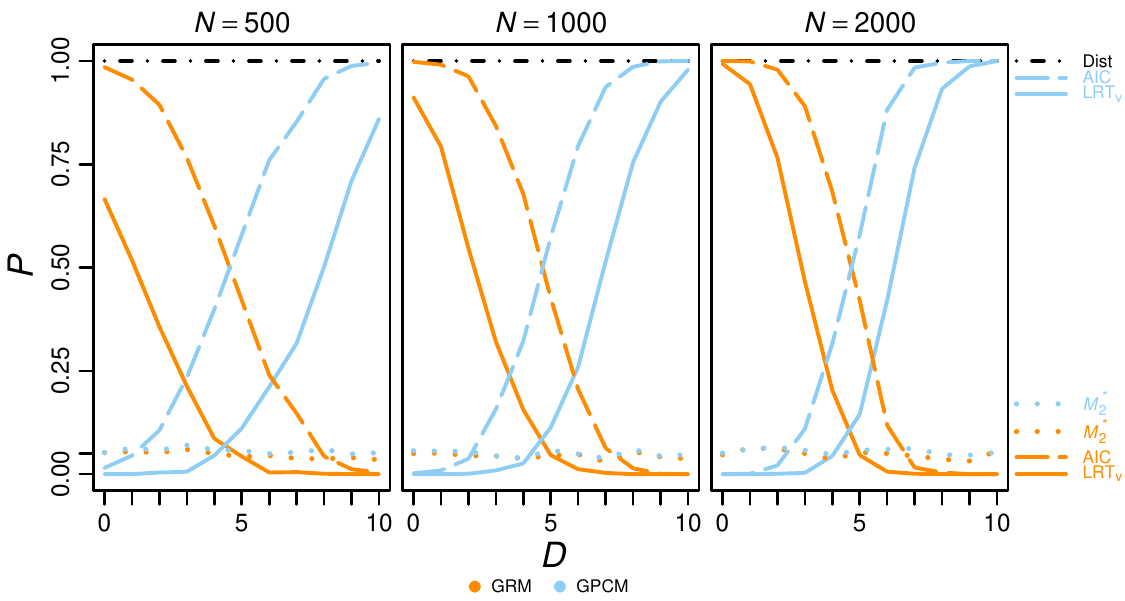}
\caption{\label{fig:sim_12}Simulation 1.2: Empirical preference/rejection rates associated with statistics. $N =$ number of persons. $J = 10$ items. $D =$ number of items of the data generating hybrid model following the GPCM.}
\end{figure}

In all conditions and all replications, the EM algorithm converged for both models. Moreover, in all conditions and all replications, second-order tests based on the condition number of the estimated information matrices of the models indicated that possible local maxima were found.

Results are displayed in Figure~\ref{fig:sim_12}, where the x-axis indicates the number of GPCM items ($D$). The three panels split the results with respect to the number of persons $N$. Within each panel, the lines represent the four statistics. For Vuong's test of distinguishability (Dist), there is only one line representing the power. For both Vuong's test of non-nested models ($\mathrm{LRT_{v}}$) and the AIC, there are two lines: one for each model representing the relative frequency of the model being preferred over the other. Note that this is symmetric for the AIC but not for the test of non-nested models due to the possibility of neither model being preferred over the other one. For the $M_2^*$ statistic, there are also two lines, one for each model, representing the relative frequency of the test statistic indicating a bad model fit.

Looking at the results for Vuong's test of distinguishability, we observe a power of one for all conditions, implying that the GRM and GPCM can be perfectly distinguished from each other in the scenarios examined. Remember that based on this test, we only conclude that the models can potentially be differentiated based on their fit. We do not, however, draw any conclusions about \emph{which} model fits the data better. While a perfect power of one may seem inordinate, note that this result was somewhat to be expected in the context of IRT modeling if and when the data generating process is informative (c.f. Simulation 1.1 and especially Simulation A.1). Contrary to other more malleable statistical models, IRT models are often more limited in their mathematical structure; specifically, their probabilistic ``predictors'' are predetermined by the model parameters themselves, resulting in predicted values that are unlikely to overlap. This effect is quite different compared to linear regression analyses, for example, where it may happen that different predictor variables make identical predictions in some populations (thereby sharing substantial overlap) which ultimately leads to competing models being statistically indistinguishable using Vuong's methodology.

Regarding Vuong's test of non-nested models, when all data generating items follow the GRM ($D = 0$), the GRM is preferred over the GPCM at a rate near $67\%$ for the condition of $N = 500$, and this rate increases up to $99\%$ as the number of persons increases. Analogously, the same pattern holds for the GPCM when all data generating items follow the GPCM ($D = 10$). Moreover, with $D$ increasing, the relative preference of the GRM over the GPCM decreases, whereas the relative preference of the GPCM over the GRM increases.

A similar pattern can be observed when inspecting the results for the AIC, although model selection based on the lowest AIC results in higher ``power'' for extreme values of $D$ (e.g., $0$ and $10$). On the other hand, this procedure does not allow for the conclusion that neither model is to be preferred, or that both models fit equally well, resulting in a relative preference rate close to chance for both models at $D = 5$. To allow for a comparison of the absolute differences in AIC values with the results reported in Simulation A.1, we again computed their mean and standard deviation, as well as their $10\%$ and $90\%$ quantiles for some selected conditions. For the condition of $N = 500$ and $D = 0$: $Mean = 18.88$, $SD = 9.73$, $Q_{10\%} = 6.40$ and $Q_{90\%} = 31.62$. For the condition of $N = 500$ and $D = 3$: $Mean = 9.87$, $SD = 7.46$, $Q_{10\%} = 1.42$ and $Q_{90\%} = 20.77$. For the condition of $N = 500$ and $D = 5$: $Mean = 8.81$, $SD = 6.91$, $Q_{10\%} = 1.43$ and $Q_{90\%} = 18.44$. Notice that with $D$ increasing (i.e., up to $D = 5$) the absolute differences in AIC values tend to become smaller, and their distribution tends to (partially) overlap with the distribution reported in Simulation A.1's Figure~\ref{fig:sim_A1}. While this is the expected behavior, this also highlights the problematic aspects of performing model selection based solely on differences in information criteria. In this case, and looking back at Simulation A.1, we are tempted to conclude that an absolute difference in AIC values of around two should not be regarded as an indication of preference of one model over the other one because we were not able to formally distinguish them based on Vuong's test of distinguishability. However, observing the results reported here, we are tempted to conclude that this difference of around two should, in fact, be regarded as an indication of preference.

Inspecting the $M_2^*$ statistic, we reject a ``good fit'' for both models at a rate close to $5\%$. We conclude that both models fit the data well, being rejected at the nominal Type I error rate independent of the data generating process. This result highlights that the $M_2^*$ statistic cannot be used for model selection as we are left with no indication of preference of one model over the other, even when all items follow either the GRM or GPCM. Admittedly, the $M_2^*$ statistic was not originally designed to be used for model selection, but rather as a statistic for evaluating the absolute fit of a model according to the first and second moment structures. Nevertheless, this simulation highlights why goodness-of-fit statistics based on a subset of the moments of the data are often insufficient for evaluating the true population generating models.

\subsubsection{Discussion}

In Simulation 1.2, we showed how the Vuong tests could be used to compare the fit of a GRM to the fit of a GPCM. To our knowledge, these are the first formal test statistics for comparing such models. We found that the two models could be reliably distinguished from one another and, in the cases of $D = 0$ and $D = 10$, Vuong's test of non-nested models was able to select the data generating model with near perfect accuracy. As seen in this simulation, applying Vuong's test of non-nested models can result in the conclusion that both models fit equally well. We argue that this is a benefit rather than a drawback, and further discuss the implications of the test of non-nested models in the General Discussion section. In the next simulation section, we apply the Vuong tests to the comparison of nested models.

\section{Simulation 2: Nested Models}

While the Vuong tests' application to non-nested IRT models is relatively novel, the statistics can also be used to test nested models. In this case, they serve as alternatives to the traditional tests based on the likelihood function, such as the traditional likelihood ratio test, Wald test, or score test \cite{eng84}. As mentioned earlier, however, the Vuong tests do not rely on the assumption that either of the competing models are correctly specified. Thus, it is reasonable to expect that the Vuong tests' properties will differ from the traditional tests in some scenarios. We study this expectation, among others, in this simulation section, and focus on nested IRT models for dichotomous data. 

In Simulation 2.1, we compare the fit of the RM to the 2PLM when the data are either generated under the RM, 2PLM, or a modified three-parameter logistic model (3PLM), relying on the latter to investigate performance under misspecification of the 2PLM. In Simulation 2.2, we compare the fit of the 2PLM to the within-item 2d-2PLM, when the data are either generated under the 2PLM or the 2d-2PLM, varying the correlation of the two latent dimensions. As already mentioned in the Introduction section, the traditional likelihood ratio test should probably not be used in this scenario, and therefore we expect inflated Type I error rates. In both simulations, we study Vuong's test of distinguishability and Vuong's test of nested models as alternatives to the traditional likelihood ratio test, and compare these further to the AIC, BIC and $M_2$ statistic (note that for dichotomous response data, $M_2 \equiv M_2^*$). Estimation defaults and assumptions were the same as in Simulation section 1, if not stated otherwise. Simulation results are reported based on the replications in which both models converged. Test statistics were evaluated at an $\alpha$ of $0.05$.

\subsection{Simulation 2.1: Comparing the RM and the 2PLM}

\subsubsection{Methods}

Simulation conditions were defined by the number of persons, $N = 500$, $1000$, or $2000$, the test length, $J = 10$, $20$, $30$, or $40$, and the data generating model either being the RM, the 2PLM or a modified 3PLM with varying lower asymptote parameters, restricted to be the same for all items. As we did not cover IRT models with lower or upper asymptotes in the Introduction section, we briefly introduce the 3PLM in this section. The 3PLM extends the 2PLM by introducing another item parameter $g_{j}$ for each item, a lower asymptote acting as a so-called ``guessing parameter'', modeling the probability of person $i$ ``solving'' item $j$ as:

\begin{equation}
  p_{ij1} = g_{j} + \frac{(1 - g_{j})}{1 + \exp(- (\beta_{j} + a_{j} \theta_{i}))}
\end{equation}

In this section, we consider a modified 3PLM, restricting these guessing parameters $g_{j}$ to be the same for all items ($g = 0.01$, $0.05$, or $0.25$) while simultaneously restricting the slopes to one for all items. Analogous to \citeA{maydeu06}, this allows us to evaluate the tests statistics' performance under misspecification of the less restrictive model, as one could argue that the 2PLM is not correctly specified when the data are generated under this modified 3PLM.

In the conditions of the RM or the modified 3PLM being the data generating model, slopes were fixed at one for all items. In the condition of the 2PLM being the data generating model, slopes were drawn from a log-normal distribution with a mean of zero and a standard deviation of $0.25$. Intercepts were drawn form the standard normal distribution.

Regarding the fitting of the RM, slopes were fixed at one for all items and the latent variance $\sigma_\theta^2$ was freely estimated. In each condition we generated 1000 datasets, and for each generated dataset, we computed six statistics after fitting the RM and 2PLM: Vuong's test of distinguishability, Vuong's test of nested models, the traditional likelihood ratio test, and each model's AIC, BIC, and $M_2$ statistic. In addition to evaluating the difference in AIC and BIC and calculating the rate of preference of the 2PLM over the RM given this difference in AIC and BIC, we checked whether Vuong's test of distinguishability, Vuong's test of nested models, and the traditional likelihood ratio test indicated preference of the 2PLM over the RM, and whether the $M_2$ statistic indicated bad model fit.

\subsubsection{Results}

\begin{landscape}
\begin{table}
\footnotesize
\centering
\addvbuffer[12pt]{
\begin{threeparttable}
\caption{\label{tab:sim_21} Simulation 2.1: Comparing the RM and the 2PLM Under Ideal Scenarios and Misspecification.}
\begin{tabular}{l *{24}{c}}
\hline
{} & {} & \multicolumn{7}{c}{$N = 500$} & {} & \multicolumn{7}{c}{$N = 1000$} & {} & \multicolumn{7}{c}{$N = 2000$} \\ \cline{3-9} \cline{11-17} \cline{19-25}
{} & {} & {} & {} & {} & {} & {} & \multicolumn{2}{c}{$M_2$} & {} & {} & {} & {} & {} & {} & \multicolumn{2}{c}{$M_2$} & {} & {} & {} & {} & {} & {} & \multicolumn{2}{c}{$M_2$}\\ \cline{8-9} \cline{16-17} \cline{24-25}
DGM & $J$ & Dist & $\mathrm{LRT_{v}}$ & $\mathrm{LRT_{t}}$ & $\mathrm{AIC}$ & $\mathrm{BIC}$ & RM & 2PLM & {} & Dist & $\mathrm{LRT_{v}}$ & $\mathrm{LRT_{t}}$ & $\mathrm{AIC}$ & $\mathrm{BIC}$ & RM & 2PLM & {} & Dist & $\mathrm{LRT_{v}}$ & $\mathrm{LRT_{t}}$ &$ \mathrm{AIC}$ & $\mathrm{BIC}$ & RM & 2PLM\\
\hline
RM & 10 & 0.04 & 0.05 & 0.06 & 0.04 & 0.00 & 0.06 & 0.05 & {} & 0.04 & 0.04 & 0.05 & 0.03 & 0.00 & 0.05 & 0.05 & {} & 0.04 & 0.04 & 0.04 & 0.03 & 0.00 & 0.05 & 0.04\\
RM & 20 & 0.02 & 0.04 & 0.06 & 0.01 & 0.00 & 0.05 & 0.05 & {} & 0.03 & 0.04 & 0.05 & 0.01 & 0.00 & 0.04 & 0.04 & {} & 0.03 & 0.04 & 0.05 & 0.00 & 0.00 & 0.05 & 0.05\\
RM & 30 & 0.01 & 0.03 & 0.04 & 0.00 & 0.00 & 0.06 & 0.07 & {} & 0.02 & 0.03 & 0.04 & 0.00 & 0.00 & 0.05 & 0.05 & {} & 0.04 & 0.04 & 0.05 & 0.00 & 0.00 & 0.05 & 0.05\\
RM & 40 & 0.01 & 0.04 & 0.06 & 0.00 & 0.00 & 0.05 & 0.05 & {} & 0.02 & 0.04 & 0.06 & 0.00 & 0.00 & 0.06 & 0.06 & {} & 0.04 & 0.06 & 0.06 & 0.00 & 0.00 & 0.04 & 0.05\\
\hline
2PLM & 10 & 0.69 & 0.75 & 0.77 & 0.73 & 0.03 & 0.49 & 0.04 & {} & 0.92 & 0.93 & 0.94 & 0.92 & 0.20 & 0.80 & 0.05 & {} & 0.99 & 0.99 & 0.99 & 0.99 & 0.58 & 0.95 & 0.05\\
2PLM & 20 & 0.96 & 0.98 & 0.98 & 0.96 & 0.03 & 0.75 & 0.06 & {} & 1.00 & 1.00 & 1.00 & 1.00 & 0.37 & 0.95 & 0.05 & {} & 1.00 & 1.00 & 1.00 & 1.00 & 0.88 & 1.00 & 0.05\\
2PLM & 30 & 1.00 & 1.00 & 1.00 & 0.99 & 0.04 & 0.83 & 0.05 & {} & 1.00 & 1.00 & 1.00 & 1.00 & 0.51 & 0.99 & 0.05 & {} & 1.00 & 1.00 & 1.00 & 1.00 & 0.97 & 1.00 & 0.05\\
2PLM & 40 & 1.00 & 1.00 & 1.00 & 1.00 & 0.04 & 0.90 & 0.07 & {} & 1.00 & 1.00 & 1.00 & 1.00 & 0.62 & 1.00 & 0.06 & {} & 1.00 & 1.00 & 1.00 & 1.00 & 0.99 & 1.00 & 0.07\\
\hline
$\mathrm{3PL_{0.01}}$ & 10 & 0.03 & 0.05 & 0.05 & 0.04 & 0.00 & 0.06 & 0.06 & {} & 0.04 & 0.05 & 0.05 & 0.04 & 0.00 & 0.07 & 0.07 & {} & 0.06 & 0.07 & 0.08 & 0.06 & 0.00 & 0.07 & 0.06\\
$\mathrm{3PL_{0.01}}$ & 20 & 0.02 & 0.05 & 0.06 & 0.01 & 0.00 & 0.05 & 0.05 & {} & 0.03 & 0.06 & 0.06 & 0.01 & 0.00 & 0.05 & 0.05 & {} & 0.05 & 0.06 & 0.07 & 0.01 & 0.00 & 0.05 & 0.05\\
$\mathrm{3PL_{0.01}}$ & 30 & 0.01 & 0.04 & 0.06 & 0.00 & 0.00 & 0.08 & 0.08 & {} & 0.03 & 0.06 & 0.08 & 0.00 & 0.00 & 0.06 & 0.06 & {} & 0.06 & 0.07 & 0.08 & 0.00 & 0.00 & 0.04 & 0.04\\
$\mathrm{3PL_{0.01}}$ & 40 & 0.01 & 0.05 & 0.06 & 0.00 & 0.00 & 0.05 & 0.06 & {} & 0.02 & 0.05 & 0.06 & 0.00 & 0.00 & 0.05 & 0.05 & {} & 0.06 & 0.07 & 0.08 & 0.00 & 0.00 & 0.07 & 0.06\\
\hline
$\mathrm{3PL_{0.05}}$ & 10 & 0.06 & 0.07 & 0.09 & 0.06 & 0.00 & 0.07 & 0.06 & {} & 0.10 & 0.11 & 0.12 & 0.10 & 0.00 & 0.08 & 0.05 & {} & 0.20 & 0.20 & 0.22 & 0.18 & 0.00 & 0.11 & 0.05\\
$\mathrm{3PL_{0.05}}$ & 20 & 0.06 & 0.10 & 0.14 & 0.03 & 0.00 & 0.08 & 0.06 & {} & 0.18 & 0.22 & 0.26 & 0.07 & 0.00 & 0.08 & 0.04 & {} & 0.41 & 0.43 & 0.45 & 0.21 & 0.00 & 0.13 & 0.05\\
$\mathrm{3PL_{0.05}}$ & 30 & 0.06 & 0.14 & 0.20 & 0.02 & 0.00 & 0.08 & 0.06 & {} & 0.24 & 0.31 & 0.36 & 0.06 & 0.00 & 0.09 & 0.05 & {} & 0.62 & 0.66 & 0.68 & 0.28 & 0.00 & 0.14 & 0.05\\
$\mathrm{3PL_{0.05}}$ & 40 & 0.05 & 0.16 & 0.23 & 0.00 & 0.00 & 0.09 & 0.07 & {} & 0.29 & 0.42 & 0.48 & 0.05 & 0.00 & 0.12 & 0.07 & {} & 0.74 & 0.78 & 0.80 & 0.32 & 0.00 & 0.17 & 0.06\\
\hline
$\mathrm{3PL_{0.25}}$ & 10 & 0.19 & 0.26 & 0.31 & 0.27 & 0.00 & 0.15 & 0.05 & {} & 0.46 & 0.53 & 0.55 & 0.50 & 0.00 & 0.30 & 0.05 & {} & 0.77 & 0.78 & 0.80 & 0.76 & 0.02 & 0.54 & 0.06\\
$\mathrm{3PL_{0.25}}$ & 20 & 0.47 & 0.61 & 0.67 & 0.40 & 0.00 & 0.27 & 0.08 & {} & 0.87 & 0.91 & 0.91 & 0.80 & 0.00 & 0.50 & 0.07 & {} & 0.99 & 1.00 & 1.00 & 0.97 & 0.05 & 0.82 & 0.08\\
$\mathrm{3PL_{0.25}}$ & 30 & 0.67 & 0.85 & 0.88 & 0.55 & 0.00 & 0.34 & 0.08 & {} & 0.98 & 0.98 & 0.98 & 0.92 & 0.00 & 0.62 & 0.10 & {} & 1.00 & 1.00 & 1.00 & 1.00 & 0.09 & 0.93 & 0.08\\
$\mathrm{3PL_{0.25}}$ & 40 & 0.78 & 0.94 & 0.96 & 0.66 & 0.00 & 0.41 & 0.15 & {} & 1.00 & 1.00 & 1.00 & 0.98 & 0.00 & 0.74 & 0.11 & {} & 1.00 & 1.00 & 1.00 & 1.00 & 0.14 & 0.96 & 0.14\\
\hline
\end{tabular}
\begin{tablenotes}[flushleft]
{\small
\textit{Note.} DGM = data generating model. $\mathrm{3PL}_{g} = $ 3PLM with lower asymptotes restricted to $g$ and slopes restricted to one for all items. $N = $ number of persons. $J = $ number of items. Values under $\mathrm{AIC}$ and $\mathrm{BIC}$ are the empirical preference rates of the 2PLM over the RM based on the differences in AIC and respectively BIC values.
}
\end{tablenotes}
\end{threeparttable}
}
\end{table}
\end{landscape}

In all conditions and all replications, the EM algorithm converged for both models. Moreover, in all conditions and all replications, second order tests based on the condition number of the estimated information matrices of the models indicated that possible local maxima were found.

Results are displayed in Table~\ref{tab:sim_21}. With the RM being the data generating model, all statistics demonstrate preference of the 2PLM over the RM or bad model fit for either model at around the nominal Type I error rate of $5\%$. However, Vuong's test of distinguishability (Dist) and the AIC and BIC generally appear to be conservative in their error control rates. When data are generated under the 2PLM, all test statistics demonstrate preference of the 2PLM over the RM with high power, increasing with the number of items and persons, and Vuong's test of nested models ($\mathrm{LRT_{v}}$) and the traditional likelihood ratio test ($\mathrm{LRT_{t}}$) show almost equivalent performance. While this also holds for the AIC, the BIC performs less well in comparison. Finally, the $M_2$ statistic is also sensitive to the 2PLM being the data generating model, indicating a bad model fit of the RM at high rates, while holding its nominal Type I error rate for the 2PLM. 

Evaluating the scenarios including misspecification, the Vuong tests and the traditional likelihood ratio test appear to be robust under minor misspecification ($g = 0.01$). However, with increasing misspecification ($g = 0.05$, or $0.25$), all tests increasingly prefer the 2PLM over the RM, a finding \citeA{maydeu06} previously reported for the traditional likelihood ratio test. Although this degree of preference of the 2PLM over the RM is generally smaller under Vuong' test of distinguishability, which showed the best performance compared to all other statistics, this difference in performance can be considered negligible in most of the scenarios examined in that the tests' performance is far from being ideal. The same conclusions hold for the traditional likelihood ratio test and the AIC and BIC as well.

\begin{figure}
\centering
\includegraphics[width=6in, height=2.4in]{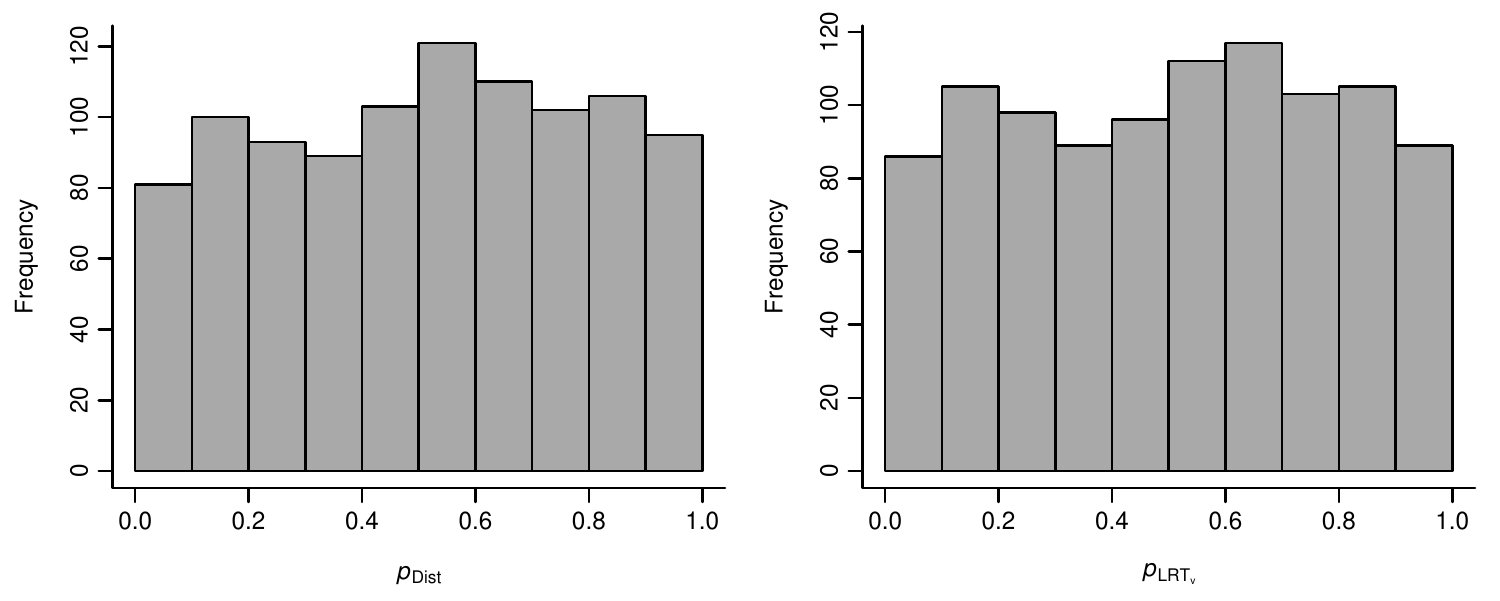}
\caption{\label{fig:sim_21} Simulation 2.1: Histograms of $p$-values for Vuong's test of distinguishability (Dist) and test of nested models ($\mathrm{LRT_{v}}$) under the null hypothesis, i.e., the RM being the data generating model. $N = 2000$ persons. $J = 10$ items.}
\end{figure}

In this simulation, we also investigated whether the empirical distributions of the Vuong test statistics' match their theoretical distributions under the null hypothesis when comparing nested models. We therefore again investigated whether the $p$-values are distributed uniformly under the null hypothesis. Figure~\ref{fig:sim_21} shows two histograms of $p$-values, one for Vuong's test of distinguishability, and one for Vuong's test of nested models for the scenario of the RM being the data generating model, $N = 2000$ and $J = 10$. Looking at these histograms, $p$-values seem to be uniformly distributed under the null hypothesis.

\subsubsection{Discussion}

In Simulation 2.1, we showed that the Vuong tests, especially Vuong's test of nested models, perform as well as the traditional likelihood ratio test when comparing nested models under ideal scenarios (i.e., the models are truly nested, the parameters to be tested lie in the interior of the parameter space, and the less restrictive model is correctly specified). However, this actually comes as no surprise, as under these conditions the equivalence of the Vuong tests and the traditional likelihood ratio test has been proven \cite<see>[Corollary 7.3, Corollary 7.5]{vuo89}. We have also seen that when the less restrictive model is severely misspecified, the Vuong tests do not necessarily perform substantially better than the traditional likelihood ratio test --- at least, in the scenarios examined here. In Simulation 2.2, we focus on nested models of different dimensions.

\subsection{Simulation 2.2: Comparing Nested Models of Different Dimensions}

\subsubsection{Method}

Simulation conditions were defined by the number of persons (fixed at $N = 2000$), the test length, $J = 10$, $20$, $30$, or $40$ and the data generating process either being the 2PLM or the within-item 2d-2PLM, varying the correlation of the two latent dimensions, $\rho = \frac{2}{3}$, $\frac{1}{3}$, or $0$. In the conditions of the 2PLM being the data generating model, person parameters were assumed to follow the standard normal distribution. In the conditions of the 2d-2PLM being the data generating model, person parameters were assumed to follow a bivariate normal distribution with means of zero and a covariance matrix with variances of one and a covariance of $\rho$. Both vectors of slopes were drawn independently from a log-normal distribution with a mean of zero and a standard deviation of $0.25$, resembling a within-item multidimensional structure with uncorrelated factor loadings, while intercepts were drawn from the standard normal distribution.

Both models were estimated via marginal ML, assuming the standard normal distribution of the person parameters for the 2PLM and a bivariate normal distribution with means of zero and the identity matrix as the covariance matrix for the 2d-2PLM. Regarding the 2d-2PLM, the second slope of the last item was always fixed at zero, resolving the rotational indeterminacy of the model. In each condition we generated 1000 datasets, and for each generated dataset we computed six statistics after fitting the models: Vuong's test of distinguishability, Vuong's test of nested models, the traditional likelihood ratio test, and each model's AIC, BIC and $M_2$ statistic. We checked whether Vuong's test of distinguishability, Vuong's test of nested models, the traditional likelihood ratio test and the difference in AIC and BIC implied preference of the 2d-2PLM over the 2PLM, and whether the $M_2$ statistic indicated bad model fit.

\subsubsection{Results}

\begin{figure}
\centering
\includegraphics[width=6in, height=2.4in]{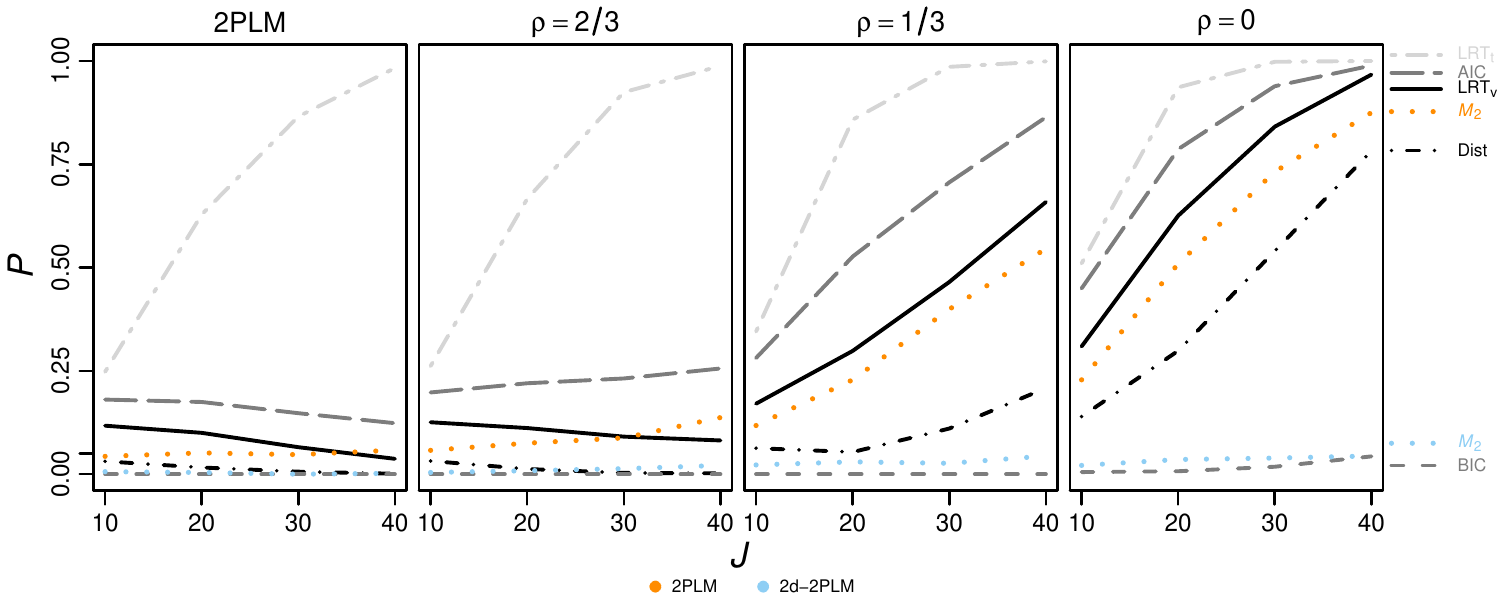}
\caption{\label{fig:sim_22}Simulation 2.2: Empirical preference/rejection rates associated with statistics. $N = 2000$ persons. $J =$ number of items. $\rho =$ correlation between the two latent dimensions under the data generating 2d-2PLM.}
\end{figure}

We observed the lowest rate of convergence of both models in the condition of $N = 2000$, $M = 20$ and $\rho = \frac{2}{3}$, where $96\%$ of the time the models successfully converged. In all conditions and all replications, second-order tests based on the condition number of the estimated information matrices of the models indicated that possible local maxima were found. 

Results are displayed in Figure~\ref{fig:sim_22}, where the x-axis shows the number of test items ($J$). The four panels split the results with respect to the data generating model being either the 2PLM or the 2d-2PLM with varying correlation $\rho$. Within each panel, the lines represent the six statistics. For Vuong's test of distinguishability (Dist), Vuong's test of nested models ($\mathrm{LRT_{v}}$), the traditional likelihood ratio test ($\mathrm{LRT_{t}}$), and the AIC and BIC, there is only one line representing Type I error rate/power. For the $M_2$ statistic, there are two lines, one for each model, representing the relative frequency of the statistic indicating bad model fit.

Regarding Vuong's test of distinguishability, Vuong's test of nested models, and the traditional likelihood ratio test, we notice that the latter test shows a highly inflated Type I error rate, implying preference of the 2d-2PLM over the 2PLM at a rate of around $25\%$, increasing with the number of items up to $98\%$ --- even though no second dimension is present in the data (the 2PLM being the data generating model). This problematic behavior of the traditional likelihood ratio test was somewhat to be expected due to a misspecification scenario being present (as described in the subsection of alternative methods and nested models and likelihood ratio tests). Importantly, however, both Vuong's test of distinguishability and test of nested models imply preference of the 2d-2PLM over the 2PLM at much more reasonable Type I error rates, with the former being slightly too conservative and the latter being more liberal. Moreover, both tests are sensitive to the correlation $\rho$ decreasing, implying increasing preference of the 2d-2PLM over the 2PLM with a peak of power for the former at around $78\%$, and $97\%$ for the latter ($J = 40$, $\rho = 0$).

Focusing on the AIC now, there is a less pronounced bias for the 2d-2PLM to be selected, implying preference of the 2d-2PLM over the 2PLM at rates of around $12\%$ to $18\%$ given no second dimension (the 2PLM being the data generating model). Analogous to Vuong's test of distinguishability and test of nested models, the AIC increasingly prefers the 2d-2PLM over the 2PLM as $\rho$ decreases. In contrast, the BIC shows itself to be strictly conservative, almost always expressing preference for the less complex 2PLM.
 
Lastly, the $M_2$ statistic implies bad model fit for the 2d-2PLM at overall Type I error rates of $1\%$ to $5\%$, being conservative when no second dimension is present (the 2PLM being the data generating model). Regarding the 2PLM, the $M_2$ statistic implies bad model fit at Type I error rates close to $5\%$, and increasingly implies bad model fit as $\rho$ decreases (up to a power of $88\%$ for $J = 40$, $\rho = 0$).

\subsubsection{Discussion}

In Simulation 2.2, we found the Vuong tests to exhibit good behavior for testing the dimension of the 2PLM. In contrast, the traditional likelihood ratio test performed quite poorly, exhibiting very large Type I error rates. Both the AIC and $M_2$ statistic exhibited reasonable performance for model selection, though we reiterate that these statistics do not provide formal tests of model comparison. In the General Discussion section, we provide further thoughts on nested models of different dimensions and future developments of the Vuong tests. In the following section, we study the Vuong test's application to IRT models using real data.

\section{Application: The Nerdy Personality Attributes Scale}

\subsection{Background}

The Nerdy Personality Attributes Scale \cite<NPAS;>{NPAS} was developed as an online questionnaire by the Open Source Psychometrics Project aiming at quantifying a ``nerdiness'' construct. The NPAS consists of 26 items in total, each rated on a five-point Likert scale, where a total of $N = 1445$ participants were collected over several months in 2015. For the purpose of this analysis we limit our demonstration to a subset of science-related items only; namely, items 1, 2, 6, 13, 22 and 23. The exact item wordings are presented in the appendix. As an example of this science-related content, item 1 states: ``I am interested in science''. We excluded $384$ participants due to failing the additional validity check items or failing to answer any of these six items. Our final dataset therefore consists of $N = 1061$ participants responding to six items.

\subsection{Method}

For this analysis we only consider the GRM and GPCM as suitable models to be fit to the data. First, we explored their fit using the AIC statistic. Second, we followed up with Vuong's test of distinguishability, and if we concluded that the models can be distinguished, we then tested which model provides the better fit using Vuong's test of non-nested models.

After having selected one of these unidimensional models, we then further wanted to test whether a two-dimensional version of the selected model provides an even better fit. Again, we first explored their overall fit using the AIC, but we also tested whether the unidimensional model fits as well as its two-dimensional version, using the traditional likelihood ratio test. We then compared these results to Vuong's test of distinguishability and Vuong's test of nested models. For all models, we also investigated absolute model fit, however, the $M_{2}^{*}$ statistic could not be computed due to too few degrees of freedom. We therefore computed the $C_{2}$ statistic instead.

\subsection{Results}

Looking at the $C_{2}$ statistic, both models fit the data well, ${C_{2}}(\mathrm{GRM})_{(9)} = 9.78$, $p = 0.369$, RMSEA $= 0.009$; ${C_{2}}(\mathrm{GPCM})_{(9)} = 10.63$, $p = 0.302$, RMSEA $= 0.013$. Examining the two models' AICs demonstrated that the GRM is preferred to the GPCM ($\mathrm{AIC}_{\mathrm{GRM}} = 17412.12$, $\mathrm{AIC}_{\mathrm{GPCM}} = 17466.53$, $\Delta_{\mathrm{AIC}} = -54.41$). Next, we followed up with Vuong's test of distinguishability and found that we could distinguish the GRM from the GPCM ($\hat{\omega}^{2}_{*} = 0.04$, $p < 0.001$). Finally, we used Vuong's test of non-nested models to compare the respective model fits. We found indeed that the GRM does fit better than the GPCM ($z = 4.41$, $p < 0.001$), and selected the GRM as the better fitting unidimensional model for these data.

Following these initial model comparisons, we were interested in whether a two-dimensional GRM provides a significantly better fit than the unidimensional model. Looking at the $C_{2}$ statistic, the two-dimensional GRM also fits the data well, ${C_{2}}(\mathrm{2d\text{-}GRM})_{(4)}= 3.73$, $p = 0.443$, RMSEA $= 0$. Examining these two models AICs', we were left with no strong evidence in favor of one model over the other ($\mathrm{AIC}_{\mathrm{GRM}} = 17412.12$, $\mathrm{AIC}_{\mathrm{2d\text{-}GRM}} = 17401.29$, $\Delta_{\mathrm{AIC}} = 10.83$). Based on the criteria of selecting the model with the lower information index, we would have chosen the 2d-GRM. Looking at the traditional likelihood ratio test, we were left with the same conclusion as well ($\chi_{(5)}^{2} = 20.83$, $p < 0.001$), and the same holds for Vuong's test of nested models ($\mathrm{LR} = 20.83$, $p = 0.022$). However, applying Vuong's test of distinguishability yielded different results: $\hat{\omega}^{2}_{*} = 0.02$, $p = 0.175$. As we have seen in Simulation 2.2, model selection based on the traditional likelihood ratio test can be misleading when comparing nested models of different dimensions, and Vuong's test of distinguishability was the only test statistic exhibiting a reasonable Type I error rate. In this scenario, Vuong's test of distinguishability is likely more reliable than the other test statistics. Therefore, based on these results we conclude that there is little reason to adopt the more complex 2d-GRM, and consequently retain the GRM as the most reasonable modeling representation for these data.

\section{General Discussion}

As described in this paper, Vuong's \citeyear{vuo89} statistical framework of model selection provides applied researchers with a useful set of statistical tests that allow for the comparison of both nested and non-nested IRT models. Central results of our simulation studies are that the tests could reliably distinguish between the GRM and GPCM, which are non-nested models whose fits are typically not formally compared. Similar results were observed when investigating the RM, 2PLM, and modified 3PLMs. Further, Vuong's tests of distinguishability and nested models generally performed as well as, or sometimes even better than, the traditional likelihood ratio test, with the latter performing poorly when comparing nested models with different numbers of latent traits, where it yielded highly inflated Type I error rates. In the discussion below, we provide some additional thoughts on indistinguishable or equally well fitting non-nested models, as well as nested models of different dimensions, and provide directions for future research. Moreover, we discuss the regularity conditions of Vuong's test statistics and address IRT models with lower and upper asymptotes.

\subsection{Non-Nested Models Being Indistinguishable or Fitting Equally Well}

As we have seen in our simulation studies, comparing non-nested models can result in Vuong's test of distinguishability concluding that two competing models are not distinguishable; in other words, the results demonstrate almost identical likelihoods for nearly all persons. Moreover, Vuong's test of non-nested models can imply that two competing models, although distinguishable, provide equal fit to the data; resulting, for instance, in the same mean log-likelihood. In Simulation A.1 we have shown that indistinguishability of non-nested models could hint at the data generating process being ``uninformative'', where neither of the competing models should be selected. In Simulation 1.2, we demonstrated that the GRM and GPCM can be distinguished when the data generating items follow either the GRM or GPCM.

For practitioners who ultimately have to choose one model, the scenario of indistinguishable or equally well fitting non-nested models is arguably harder than the nested case. If the two competing models are indistinguishable or fit equally well and differ in their number of model parameters, practitioners can argue for the merits of the less complex model, following the principle of parsimony, as is common when comparing nested models. If the two competing models share the same number of model parameters (e.g., the GRM and the GPCM) and the Vuong tests suggest that the models are either indistinguishable or fit equally well, we argue that based on statistical information alone there is no justification for choosing either model. In this sense,  additional data is required before explicit support for either competing model can be reached. 

Situations arise, however, where practitioners will want to select one model for further analysis purposes. In this scenario, one may argue that not much insight is gained from the Vuong tests. However, we argue that the Vuong tests provide some additional insight that can be achieved. For instance, after inspecting the sign and value of the test statistic of Vuong's test of non-nested models, practitioners can gain a descriptive index similar to information criteria, which may be more natural to interpret because it can be rescaled to the $Z$ scale. Moreover, this scenario may also allow practitioners to revisit their theoretical justification for either competing model. In our opinion, this is a benefit rather than a drawback of these tests, particularly when compared to model selection solely based on differences in information criteria, whereby practitioners will often interpret even small differences as an indication of preference for one model over another \cite<see, e.g.,>{stochl13}. Nonetheless, we do not want to undermine the high practical use of information criteria for model selection. As demonstrated in the Application section, we believe that the combination of evaluating differences in information criteria and applying the Vuong tests, as well as possibly other statistical tests, allows practitioners to select one model for their further analysis with greater degrees of confidence.

\subsection{Nested Models of Different Dimension}

In our simulation studies, using the traditional likelihood ratio test for testing nested models of different dimensions (e.g., testing the 2PLM vs. 2d-2PLM) resulted in highly inflated Type I error rates. As outlined in the Introduction section, one could argue that this is due to a a misspecification scenario in combination with the parametrization of (exploratory) multidimensional IRT models (i.e., assuming the identity matrix as the covariance matrix of the latent traits). To our knowledge, the magnitude of the severity for the traditional likelihood ratio test has not been strongly emphasized in the literature of IRT.

In practice, it may be a possible solution to implement a bootstrap methodology \cite{efron98} to better approximate the distribution of the traditional likelihood ratio test in the scenarios described in this paper. For example, in the context of finite mixture models, boostrapping the traditional likelihood ratio test has been proven to be quite successful \cite<see, e.g.,>{mclachlan87,feng96}. However, in the scenarios we examined, we demonstrated that the Vuong tests (especially Vuong's test of distinguishability) are robust alternatives to the traditional likelihood ratio test, holding more reasonable Type I error rates, while also demonstrating reasonable power. Nevertheless, we encourage future research to theoretically investigate the problem of the traditional likelihood ratio test in the context of nested IRT models with different dimensions. As an example of alternative approaches in the context of linear mixed models with one variance component, \citeA{crainiceanu04} were able to theoretically derive the non-standard finite sample and asymptotic distribution of the traditional likelihood ratio test.

\subsection{Regularity Conditions and Models with Lower and Upper Asymptotes}

As stated by \citeA{vuo89}, and also discussed in \citeA{meryou16}, the conditions under which the assumptions of the Vuong tests hold are quite general (e.g., existence of second-order derivatives of the log-likelihood, invertibility of the models' information matrices, and i.i.d distributed data vectors). As discussed in \citeA{jeffries03} and \citeA{wilson15}, applying Vuong's tests to compare mixture models with different number of components can violate the invertibility requirement due to the lower dimensional model lying on the boundary of the parameter space of the higher dimensional model, which can result in inflated Type I error rates. In the context of IRT, researchers are familiar with models including lower and upper asymptotes, which may share similar limitations. Comparing the 2PLM to the 3PLM, for example, mimics the same problems as described above due to the 2PLM lying on the boundary of the parameter space of the 3PLM, restricting all guessing parameters to zero. \citeA{brown15} point out that in this scenario, the application of the traditional likelihood ratio test results in deflated Type I error rates, while \citeA{cha17} suggest similar issues when computing likelihood-based confidence intervals for these types of models. As such, similar problems may arise when applying Vuong's tests. Therefore, although technically already possible, we do not wish to encourage researchers and practitioners to compare models including lower and upper asymptotes until future research has systematically examined these scenarios both theoretically and by simulation studies.

\subsection{Conclusion}

Vuong's \citeyear{vuo89} tests provide researchers and practitioners with effective methods for comparing the fits of both nested and non-nested IRT models. We have shown in this paper that the statistics generally exhibit desirable properties, especially compared to statistics that are traditionally used for model comparison in IRT. Overall, we believe that the Vuong tests, in combination with other model selection procedures (such as information criteria), allow for performing model selection with high confidence for the IRT modeling applications studied herein. While computation and evaluation of the Vuong tests is generally difficult, the implementations in the R packages {\em mirt} \cite{mirt} and {\em nonnest2} \cite{nonnest2} make the statistics generally accessible to applied researchers and practitioners. We look forward to future extensions of the statistics to boundary scenarios (e.g., the 3PLM) and to non-traditional IRT models, such as the explanatory item response framework described by \citeA{debwil04}.

\section{Computational Details}

All results were obtained using the R system for statistical computing \cite{r18} version 3.5.1, employing the add-on packages {\em MASS} \cite{MASS} version 7.3-51.1 for simulating person parameters from a bivariate normal distribution, {\em mirt} \cite{mirt} version 1.29 for simulating data, fitting of the models and information matrix, log-likelihood derivatives, traditional likelihood ratio test, AIC, BIC and $M_2$/$M_2^*$/$C_2$ computation, {\em nonnest2} \cite{nonnest2} version 0.5-2 for carrying out the Vuong tests, and {\em SimDesign} \cite{SimDesign} version 1.13 for carrying out the simulation studies. R and the packages {\em MASS}, {\em mirt}, {\em nonnest2} and {\em SimDesign} are freely available under the General Public License from the Comprehensive R Archive Network at \url{https://cran.r-project.org/}. Numerical values were rounded based on the IEC 60559 standard. Code for replicating our results is available at \url{https://github.com/sumny/vuong\_mirt\_code}. %Code for replicating our results can be obtained from the first author.

\bibliography{refs}
\appendix

\section{Simulation A.1: Comparing the GRM and GPCM When Data Follow a Binomial Distribution}

\subsection{Method}

Simulation conditions were defined by the number of persons, $N = 500$, $1000$, or $2000$, and the length of the test (fixed at $J = 10$). In each condition, 1000 datasets were generated from a binomial distribution with hyper-parameters $n = 3$ (number of trials) and $p = 0.5$ (success probability for each trial) by drawing $N \cdot J$ values using the {\em rbinom} function to fill a $N \times J$ item response matrix of values between zero and three. This model serves as a generalization of the data generating process investigated by \citeA{wood78} to polytomous data. Note that contrary to Simulation 1.1, the data generating process used here does not necessarily imply that the null hypothesis of Vuong's test of distinguishability holds. Nevertheless, it is a priori reasonable to assume that the test results should not demonstrate any systematic preferences for or against a given model.

In each condition, and for each generated dataset we computed four statistics after fitting the models: Vuong's test of distinguishability, Vuong's test of non-nested models, and each model's AIC and $M_2^*$ statistic. We checked whether the models could be distinguished, and if this was the case, whether the non-nested test implied preference of one model over the other one. We further checked which model was to be preferred based on the lower AIC, and whether the $M_2^*$ statistic indicated bad model fit.

\subsection{Results}

\begin{table}
\footnotesize
\centering
\addvbuffer[12pt]{
\begin{threeparttable}
\caption{\label{tab:sim_A1} Simulation A.1: Comparing the GRM and the GPCM When Data Follow a Binomial Distribution.}
\begin{tabular}{l c c c c c c c c c c}
\hline
{} & \multicolumn{10}{c}{Empirical Preference/Rejection Rates} \\ \cline{2-11}
{} & {} & \multicolumn{4}{c}{GRM} & {} & \multicolumn{4}{c}{GPCM} \\ \cline{3-6} \cline{8-11}
{} & {} & \multicolumn{2}{c}{$\mathrm{LRT_{v}}$} & {} & {} & {} & \multicolumn{2}{c}{$\mathrm{LRT_{v}}$} & {} & {} \\ \cline{3-4} \cline{8-9}
$N$ & Dist & all & (Dist sgn.) & AIC & $M_2^*$ & {} & all & (Dist sgn.) & AIC & $M_2^*$ \\
\hline
500 & 0.00 & 0.01 & (0.00) & 0.55 & 0.01 &  & 0.00 & (0.00) & 0.45 & 0.01 \\
1000 & 0.01 & 0.01 & (0.00) & 0.53 & 0.01 &  & 0.00 & (0.00) & 0.47 & 0.01 \\
2000 & 0.01 & 0.00 & (0.11) & 0.49 & 0.01 &  & 0.01 & (0.00) & 0.51 & 0.02 \\
\hline
\end{tabular}
\begin{tablenotes}[flushleft]
{\small
\textit{Note.} all = using all replications for checking the preference of the non-nested Vuong test ($\mathrm{LRT_{v}}$). Dist sgn. = using only the replications in which Vuong's test of distinguishability (Dist) yielded significant results. $N = $ number of persons. $J = 10$ items.
}
\end{tablenotes}
\end{threeparttable}
}
\end{table}

\begin{figure}
\centering
\includegraphics[width=4.5in, height=2.4in]{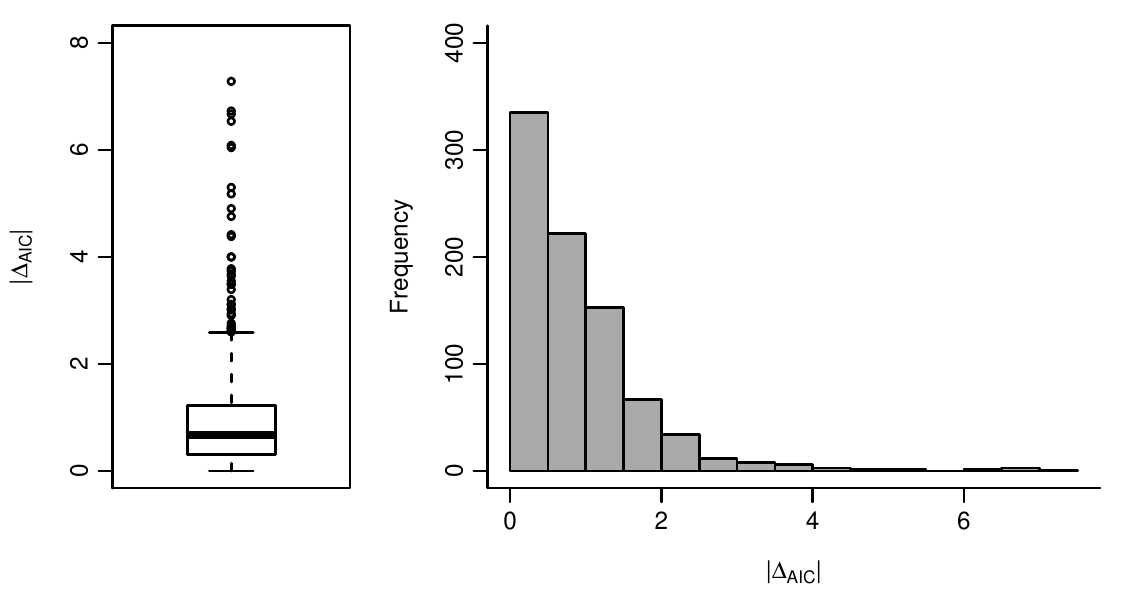}
\caption{\label{fig:sim_A1} Simulation A.1: Boxplot and histogram of the absolute differences in AIC values for the GRM and GPCM. $N = 500$ persons. $J = 10$ items.}
\end{figure}

In the condition of $N = 500$, the EM algorithm converged for both models in $85\%$ of the replications. For the condition of $N = 1000$, this was the case in $97\%$ of the replications and in the condition of $N = 2000$, this was the case in $100\%$ of the replications. In all conditions and all replications, second-order tests based on the condition number of the estimated information matrices of the models indicated that possible local maxima were found. 

Table~\ref{tab:sim_A1} summarizes the simulation results. Regardless of the number of persons, Vuong's test of distinguishability (Dist) indicates at a rate of $1\%$ that the GRM and GPCM can be distinguished. Recall that this statistical test is not designed to determine which of the two competing models provides the better fit to the data. At a rate of around $1\%$, Vuong's test of non-nested models ($\mathrm{LRT_{v}}$) prefers the GRM over the GPCM, and at a rate of around $1\%$ the GPCM is to be preferred over the GRM. However, recall that Vuong's test of non-nested models can only be applied validly if the test of distinguishability yielded a significant result beforehand. Looking only at these few replications, the GRM is to be preferred over the GPCM at a rate of $0\%$ to $11\%$, and the GPCM is to be preferred over the GRM at a rate of $0\%$. As to be expected, performing model selection based on the lower AIC results in choosing either model at a rate of $50\%$. Finally, the $M_2^*$ statistic indicates bad model fit for both models at a maximum rate of $2\%$.

Figure~\ref{fig:sim_A1} shows both a boxplot as well as a histogram of the absolute differences in AIC values for the the condition of $N = 500$. While these absolute differences tend to be small ($Mean = 0.91$, $SD = 0.91$, $Q_{10\%} = 0.13$ and $Q_{90\%} = 1.86$), substantial differences do occur nevertheless, making model selection based on the lower AIC quite misleading in some scenarios. These are potentially misleading because a researcher may conclude that one model is notably more supported by the data than another, when in fact the fit to the data is based completely on noise variation.

\subsection{Discussion}

In Simulation A.1, we showed that Vuong's test of distinguishability is useful in the context of IRT modeling, when the data generating process is uninformative for both competing models, i.e., when the data follow a binomial distribution. In this scenario, there is no basis for asking the question whether the GRM or the GPCM provides the better fit to the data, and the results from this test of distinguishability tells us exactly this; i.e., that the models result in nearly identical likelihoods for all persons. In contrary, comparing these two competing models based on their AIC can lead to misleading conclusions: In the scenarios investigated, selecting the model with the lower AIC results in falsely declaring one model as the ``better fitting'' one simply by chance. Relying on cut-off heuristic values, such as interpreting an absolute difference in AIC larger than ten as ``substantial'', can mitigate this problem to some extent; however, due to the arbitrariness being involved in declaring such a cut-off value and other factors, such as sample size variability of information criteria, this procedure still leaves plenty of room for false positive declarations of one model advertised as the ``better fitting'' one. 

\section{The Nerdy Personality Attributes Scale (NPAS)}

In the Application section, we study the Vuong tests' performance using six items of the NPAS \cite{NPAS}, rated on a five-point Likert scale ($0 =$ Disagree, $2 =$ Neutral and $4 = $ Agree). In this appendix, we provide the wording of these six items:

\begin{enumerate}[align=parleft]

\item[Q1]{I am interested in science.}
\item[Q2]{I was in advanced classes.}
\item[Q6]{I prefer academic success to social success.}
\item[Q13]{I would describe my smarts as bookish.}
\item[Q22]{I enjoy learning more than I need to.}
\item[Q23]{I get excited about my ideas and research.}

\end{enumerate}

\end{document}